\documentclass[aps,prb,reprint,longbibliography]{revtex4-2}

\usepackage{graphicx}
\usepackage{siunitx}
\usepackage{bbm}
\usepackage{amsmath,amsfonts,amssymb,braket,bbold,mathtools,array}
\usepackage[colorlinks,citecolor=blue,linkcolor=blue,urlcolor=blue,bookmarks=false,hypertexnames=true]{hyperref}

\newcommand{\eps}{\varepsilon}

\newcommand{\ain}{a_\mathrm{in}}
\newcommand{\aout}{a_\mathrm{out}}

\begin{document}

\title{Interplay of Pauli blockade with electron-photon coupling in quantum dots}

\author{Florian Ginzel}
\author{Guido Burkard}
\affiliation{Department of Physics, University of Konstanz, D-78457 Konstanz, Germany}


\begin{abstract}
Both quantum transport measurements in the Pauli blockade regime and microwave cavity transmission measurements are important tools for spin-qubit readout and characterization. Based on a generalized input-output theory we derive a theoretical framework to investigate how a double quantum dot (DQD) in a transport setup interacts with a coupled microwave resonator while the current through the DQD is rectified by Pauli blockade. We show that the output field of the resonator can be used to infer the leakage current and thus obtain insight into the blockade mechanisms. In the case of a silicon DQD, we show how the  valley quasi-degeneracy can impose limitations on this scheme. We also demonstrate that a large number of unknown DQD parameters including (but not limited to) the valley splitting can be estimated from the resonator response simultaneous to a transport experiment, providing more detailed knowledge about the microscopic environment of the DQD. Furthermore, we describe and quantify a back-action of the resonator photons on the steady state leakage current.
\end{abstract}

\maketitle

\section{Introduction\label{sec_intro}}

Spin qubits in few-electron quantum dots (QDs) \cite{RMP-spinQubits} are advancing to become one of the leading platforms for quantum information, with numerous demonstrations of high-fidelity quantum operations on the few-qubit scale \cite{Xue2021,Mills2021,Noiri2022,Mills2022}. Great hopes to achieve the required scalability lie with circuit quantum electrodynamics (cQED) implementations \cite{CQED_review}, relying on the dipole moment of electrons in a double QD (DQD). After strong  spin-photon coupling  \cite{PhysRevB.86.035314,StrongCouplingPrinceton,Samkharadze1123,Landig2018}
and cavity-mediated spin-spin coupling
\cite{Borjans2020,Harvey-Collard2022}
have been demonstrated,
photon-mediated two-qubits gates \cite{PhysRevA.69.042302,PhysRevB.74.041307,PhysRevB.97.235409,Benito2019} and resonator-based spin readout \cite{Peterson2010,PhysRevLett.110.046805,House2015,PhysRevB.100.245427,Crippa2019,Vandersypen_readout,PhysRevX.10.041010} are conceivable.

Another well established instrument in the spin qubit-toolbox is Pauli blockade \cite{RevModPhys.79.1217,RMP-spinQubits}. A DQD with two electrons can be in a state with one electron in each dot or one doubly occupied QD. Due to the Pauli exclusion principle not all two-electron states are allowed in a doubly occupied QD. For example, two electrons forming a spin triplet in two dots cannot be merged into one QD unless the excited orbital state becomes available \cite{Fujisawa2002}. In a closed system the Pauli blockade can be harnessed for initialization \cite{doi:10.1126/science.1116955,Maune2012,PhysRevApplied.10.054026} and readout \cite{doi:10.1126/science.1116955,PhysRevLett.103.160503} of different types of spin qubits.

In a transport setup Pauli blockade can lead to a rectification of the current through a DQD already occupied with one electron \cite{doi:10.1126/science.1070958}. The blockade can be partially lifted by interactions that mix the spin states such as spin-orbit interaction \cite{PhysRevB.80.041301,PhysRevB.81.201305}, hyperfine interaction \cite{doi:10.1126/science.1113719,Johnson2005,PhysRevLett.96.176804} or cotunneling processes \cite{PhysRevB.84.245407,Lai-SciRep-2011}. The observation of Pauli blockade is proof of a large single-dot singlet-triplet splitting and therefore a crucial first step towards spin qubit applications.

Pauli blockade has already been studied in the presence of a resonant drive which can cause transitions between the spin-like eigenstates of the DQD, thus significantly altering the line shape of the leakage current~\cite{S_nchez_2008,Hao2014,PhysRevB.104.085421}. This can be utilized to gain information about the DQD spectrum \cite{PhysRevLett.108.166801}. An off-resonant probe field which does not lift the blockade can dispersively interact with the dipole moment that goes along with the DC current. It has been demonstrated in GaAs QDs that this dipole moment can be harnessed to detect the lifting of the blockade in a cavity transmission measurement \cite{doi:10.1021/acs.nanolett.5b01306,PhysRevLett.122.213601}.

Silicon is one of the most promising host materials for spin qubits as it allows for  long spin coherence times \cite{Zwanenburg2013}. In QDs based on silicon, a valley pseudospin arises from the degenerate conduction band minima \cite{RevModPhys.54.437,RMP-spinQubits}. It is known from  carbon-based spin qubits \cite{Trauzettel2007,PhysRevB.88.125422} that the valley degree of freedom makes Pauli blockade much more intricate and subtle \cite{PhysRevB.80.201404,PhysRevB.82.155424,PhysRevB.82.155312} since it allows spin triplets in the orbital ground state of a doubly occupied QD, as long as the total wavefunction is antisymmetric under particle exchange.
Furthermore, in silicon, the parameters of the valley Hamiltonian are largely determined by the microscopic environment of the QDs \cite{PhysRevB.80.081305,PhysRevB.84.155320,PhysRevResearch.2.043180,Wuetz2021} with only limited possibilities to address them experimentally after the fabrication process \cite{PhysRevResearch.2.043180,PhysRevApplied.13.034068,Yang2013,doi:10.1063/1.4972514}. Microwave resonators have proven useful, here, to measure valley splitting and the inter-valley tunneling matrix elements in a given device \cite{PhysRevB.94.195305,PhysRevLett.119.176803,Russ_2020,Borjans2021}.

In this article, a comprehensive theory for electronic transport through a DQD simultaneously coupled to a microwave resonator is developed. We quantitatively investigate the resonator response to the leakage current in the Pauli spin blockade regime and describe an additional enhancement or suppression of the leakage current due to the electron-photon coupling. The analysis is extended to the case of a Pauli blockade including valley degree of freedom, where we discuss potential complications and present a scheme for a resonator-aided measurement of the leakage current. In particular, we discuss the yet largely unexplored regime of the dispersive interaction between the current and the probe field where the resonator transmission is sensitive to the spin and valley physics of the DQD with only a minimal back-action. This regime holds the promise of a reduction of complexity in large-scale QD spin qubit devices because the number of specialized methods and detectors could potentially be reduced by replacing them by a single resonator.

The remainder of this article is organized as follows. In Sec.~\ref{sec_model} we introduce the theoretical framework of the analysis. In Sec.~\ref{sec_discussion} we discuss the interaction of spin blockade and the resonator coupling. In Sec.~\ref{sec_valley} we include a lifted valley degeneracy into the discussion and present schemes for a resonator-aided current measurement in this case as well as for the measurement of unknown DQD parameters. Finally, in Sec.~\ref{sec_summary} the results are summarized.

\section{Model\label{sec_model}}

In this section we develop a model for a DQD which is shunted in series and coupled to electronic reservoirs as well as a microwave resonator. In Sec.~\ref{sec_Hamiltonian} the Hamiltonian is introduced and in Sec.~\ref{sec_IOTheory} we adopt a generalized input-output (IO) theory \cite{PhysRevB.101.155406} and develop a treatment of the Pauli blockade beyond the original generalized IO formalism.

\subsection{Hamiltonian\label{sec_Hamiltonian}}

\begin{figure}
\begin{center}
\includegraphics[width=0.5\textwidth]{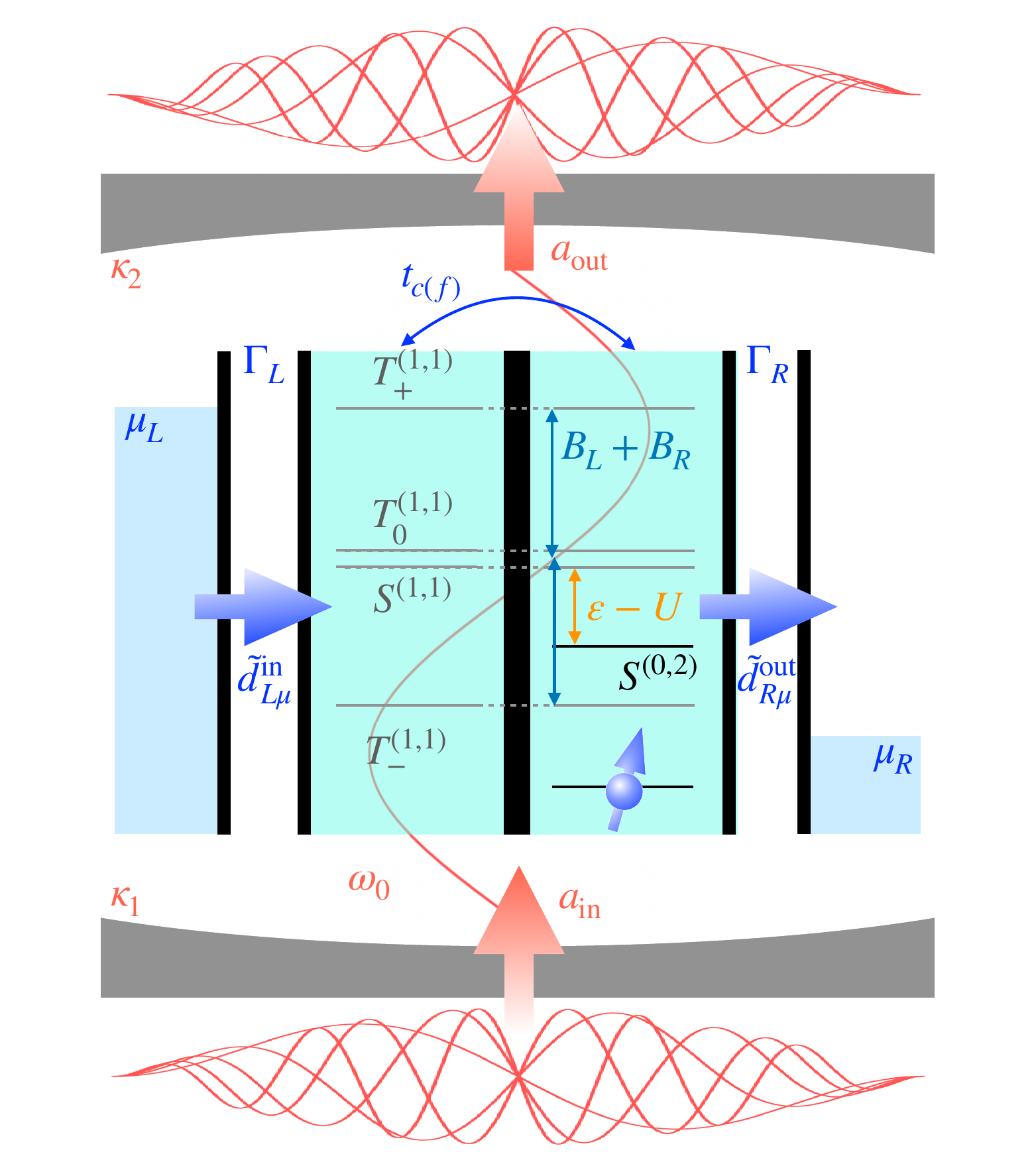}
\caption{Schematic rendering of the DQD energy levels (without valley degree of freedom for simplicity). The DQD is coupled to source (left, $L$) and drain (right, $R$) leads as well as a resonator with two ports. The spin degeneracy is lifted by the Zeeman splitting $B_{L(R)}$. The onsite potentials $V_j$ of the QDs are tuned such that the right dot is always occupied by one electron. A second electron can tunnel from the source to the left QD, between the QDs via the spin conserving (flipping) tunneling $t_{c(f)}$ and from the right QD to the drain, resulting in an electric current $I$ though the system. For $\eps - U = V_L - V_R - (U_R-U_{LR})=0$ the tunneling between the singlets is elastic. The electric dipole moment of the moving electrons couples to the electric field inside a cavity field with resonance frequency $\omega_0$. If the spin structure of a two-electron state is incompatible with a doubly occupied QD, Pauli blockade occurs and the current is suppressed. This is the case for the spin triplet states $T_{0(\pm)}$. The interaction with the environment is described by the input and output fields along with the coupling rates $\Gamma_{L(R)}$, $\kappa_{1(2)}$.\label{fig_systemsketch}}
\end{center}
\end{figure}

To describe all relevant interactions as depicted in Fig.~\ref{fig_systemsketch} we introduce the Hamiltonian $H=H_S + H_E + H_I$, where the system Hamiltonian $H_S=H_\mathrm{QD} + H_\mathrm{res} + H_\mathrm{dip}$ contains the DQD, the microwave resonator $H_\mathrm{res}$ and the dipole interaction $H_\mathrm{dip}$. The environment $H_E$ comprises the source and drain leads of the DQD and the photonic reservoirs. The interaction $H_I$ between the system and the environment will later be captured by the generalized IO theory.

To model a double quantum dot where only the lowest orbital state is available we use an extended Hubbard Hamiltonian \cite{RevModPhys.79.1217,RMP-spinQubits},
\begin{eqnarray}
    H_\mathrm{QD} &=& \sum_{j,\sigma,v} E'_{j\sigma v} d_{j\sigma v}^\dagger d_{j\sigma v} + \sum_{\sigma, \sigma',v,v'} \left( t_{\sigma v \sigma' v'} d_{R\sigma'v'}^\dagger d_{L\sigma v} \right. \nonumber \\
    && \left. + \mathrm{h.c.} \right) + U_{LR} n_L n_R + \sum_j \frac{U_j}{2} n_j \left( n_j - 1\right).
\end{eqnarray}
Here, $d_{j\sigma v}^{(\dagger)}$ annihilates (creates) an electron with spin $\sigma = \uparrow,\downarrow$ and valley index $v=\pm$ in QD $j=L,R$ and $n_j = \sum_{\sigma,v} d_{j\sigma v}^\dagger d_{j\sigma v}$ denotes the occupation number operator of dot $j$. We have introduced the onsite energies $E'_{j\sigma v} = V_j + B_j (\sigma_z)_{\sigma\sigma}/2 + \Delta_j (\sigma_z)_{vv}$ with the electric potential $V_j$, Zeeman splitting $B_j$ and valley splitting $\Delta_j$. Here, $\sigma_z$ denotes the Pauli $z$ matrix. Inter-dot tunneling ($t_{\sigma v \sigma' v'}$) can either be spin conserving, $t_c$, or spin flipping, $t_f$, and we define the valley phase difference $\varphi_v$ such that $t_{c(f)}\cos\varphi_v$ is the matrix element for intra-valley tunneling and $t_{c(f)}\sin\varphi_v$ is the matrix element for inter-valley tunneling. The Coulomb repulsion between electrons in adjacent QDs (the same QD) is given by $U_{LR}$ ($U_j$, $j=L,R$).

The resonator is modeled as a single-mode harmonic oscillator, $H_\mathrm{res} = \omega_0 a^\dagger a$ with resonance frequency $\omega_0$ and ladder operator $a$. The interaction between the DQD and the resonator photons is given by \cite{Cohen-Tannoudji,ScullyZubairy}
\begin{equation}
    H_\mathrm{dip} = \frac{g_0}{2} \left( a^\dagger +a \right) \sum_{\sigma, v}\left( d_{L\sigma v}^\dagger d_{L\sigma v} -d_{R\sigma v}^\dagger d_{R\sigma v} \right).
\end{equation}

We assume that each QD $j$ is coupled to one fermionic reservoir where $c_{kj\sigma v}^{(\dagger)}$ annihilates (creates) an electron with wavenumber $k$, spin $\sigma$ and valley $v$. At the same time, the resonator interacts with one photonic reservoir with ladder operator $b_{x,\omega}$ at each port $x=1,2$,
\begin{equation}
    H_E = \sum_{j,k,\sigma,v} \varepsilon_{kj\sigma v} c_{kj\sigma v}^\dagger c_{kj\sigma v} + \sum_x \int \mathrm{d} \omega\, \omega b_{x,\omega}^\dagger b_{x,\omega} .
\end{equation}
The electrons can tunnel between reservoir $j$ and QD $j$ with a matrix element $\tau_{kj\sigma v}$ and for each port of the resonator we define the coupling $g_x(\omega)$ to the continuum. The interaction of the electrons and the cavity photons with their respective reservoirs is described by the Hamiltonian term,
\begin{eqnarray}
    H_I &=& \sum_{j,k,\sigma,v} \tau_{kj\sigma v} \left(c_{kj\sigma v}^\dagger d_{j\sigma v} + d_{j\sigma v}^\dagger c_{kj\sigma v} \right) \nonumber\\
    &+& \sum_x \int \mathrm{d}\omega \, g_x(\omega) \left( b_{x,\omega}^\dagger a + a^\dagger b_{x,\omega}\right) .
\end{eqnarray}
For the remainder of the derivation all operators are treated as time-dependent.

To investigate Pauli blockade we restrict the system to be close to the triple point of the charge configurations $(n_L,n_R) = (1,1)$, $(0,2)$, $(0,1)$ where $\eps = V_L -V_R$ is given by $\eps = U$, with $U=U_R-U_{LR}$. Under these premises the two-electron Hilbert space of the DQD is spanned by ten supertriplet states and six supersinglets with the (1,1) charge configuration and six supersinglets with the (0,2) charge configuration \cite{PhysRevB.82.155424,Rohling_2012}. In the basis of the charge degree of freedom we separate the two-electron states into the (1,1) states $|1i\rangle$, $i=1,...16$, and the (0,2) states $|2i\rangle$, $i=1,...6$. With non-zero interdot hopping these states hybridize and form the eigenstates $|\mu\rangle = \sum_{j,j',\sigma, \sigma', v, v'} \alpha^{(\mu)}_{j\sigma v.j'\sigma'v'} d^\dagger_{j'\sigma'v'} d^\dagger_{j\sigma v} |\mathrm{vac}\rangle$ of the Hamiltonian $H_\mathrm{QD}$, $\mu = 1,...22$, where $|\mathrm{vac}\rangle$ is the vacuum state and $\alpha^{(\mu)}$ are the coefficients of the basis change,
and where the corresponding eigenenergies are denoted $E_\mu$. If the valley degree of freedom is disregarded, the (1,1) states form a spin triplet and one singlet and there is one (0,2) singlet.

The restriction to the triple point of $(1,1)$, $(0,2)$, and $(0,1)$ allows a change of the particle number of the DQD only in processes that add the second electron to a single-occupied DQD or remove one electron from a double-occupied DQD. With this restriction of the electron number we can define $d_\mu^{(\dagger)}$ that annihilates (creates) the two-electron eigenstate $\mu$ out of the vacuum~\cite{AQMbook}. The operators $d_\mu^\dagger d_\mu$ yield the probability of finding the system in the eigenstate $\mu$, which we will use for the derivation of the leakage current. It is important to note that $d_\mu^{(\dagger)}$ are not multi-particle creators and annihilators on the entire Fock space; they are only defined for up to two particles.

\subsection{Input-Output Theory\label{sec_IOTheory}}

Input-output (IO) theory is a powerful tool for the modelling of cavity-coupled qubits \cite{PhysRevA.30.1386,PhysRevB.94.195305,CQED_review}. Here, we apply a generalized version to combine the treatment of the electronic transport process and the description of the resonator field in one formalism \cite{PhysRevB.101.155406}.

First, the Hamiltonian is transformed into a rotating frame, $\tilde H = U_r H U_r^\dagger+ i (\frac{\partial}{\partial t} U_r) U_r^\dagger$, denoted by the tilde and defined by
\begin{equation}
    U_r = \exp(-i \omega_p t a^\dagger a)\exp \left(-i t \sum_\mu \Omega_\mu d_\mu^\dagger d_\mu \right),
\end{equation}
with the frequency $\omega_p$ of the probe field that is injected into the resonator. With more than two electronic levels, finding a rotating frame that removes all time-dependence from the Hamiltonian is possible only in special cases. In general, a rotating wave approximation (RWA) could lead to the (unintended) negligence of certain transitions. To account for cases where a RWA is inappropriate we leave $\Omega_\mu$ general and will give both a solution with RWA and beyond the RWA.

In the system under consideration a RWA can be made for $t_f/t_c, \tan(\varphi_v) \ll 1$, and $B_L/B_R\approx 1 \approx \Delta_L/\Delta_R$. This is because the two-particle states $|\tilde\mu\rangle$ approximately arrange themselves in six pairs of bonding and antibonding molecular supersinglet states (only one pair of singlet states in the absence of the valley pseudospin) $\tilde\mu = s_1,...s_6$ and $s'_1,...s'_6$ and a set of ten (three) decoupled supertriplet states~\cite{PhysRevA.102.063113,PhysRevB.80.041301}, and $H_\mathrm{dip}$ couples only within the pairs~\cite{PhysRevB.104.085421,PhysRevLett.122.213601,PhysRevB.101.155406}. The choice $\Omega_{s'_i}=\omega_p$, $i=1,...6$, and all other $\Omega_\mu =0$ approximately then removes time dependence. If our theory is applied to other Hamiltonians, a case-by-case analysis is required which rotating frame can be used and whether a RWA is useful or not. It is also possible to choose $\Omega_\mu = 0$ and thus use no rotating frame.

Following the procedures of IO theory \cite{PhysRevA.30.1386,PhysRevB.101.155406}, the Heisenberg equations of motion for the reservoir operators are formally integrated to eliminate the reservoir operators in the equations of motion of the system operators. This results in the Langevin equations,
\begin{eqnarray}
    \frac{\mathrm{d}}{\mathrm{d}t} \tilde a &=& -(i \Delta_c + \kappa/2) \tilde a + \sum_{x=1,2} \sqrt{\kappa_x} \tilde a_{\mathrm{in},x} \nonumber\\
    && - i \sum_{\mu,\nu} g_{\mu\nu} e^{-i(\omega_p - \Omega_\nu + \Omega_\mu)(t-t_0)} \tilde{d}_\nu^\dagger \tilde{d}_\mu , \\
    \frac{\mathrm{d}}{\mathrm{d}t} \tilde d_\mu &=& -\left[i(E_\mu - \Omega_\mu) +\sum_{j=L,R} \left(1- n_F^j(E_\mu)\right) \tilde \Gamma_{j\mu} \right] \tilde d_\mu \nonumber\\
    && - i \sum_\nu \left( g_{\nu\mu} a e^{i(\omega_p - \Omega_\mu + \Omega_\nu)(t-t_0)} + \mathrm{h.c.} \right) \tilde d_\nu \nonumber\\
    &&+ \sqrt{2\pi} \left( \tilde d_{L\mu}^\mathrm{in}(t) + \tilde d_{R\mu}^\mathrm{in}(t) \right)\left(1- \sum_\nu \tilde d_\nu^\dagger \tilde d_\nu \right).
\end{eqnarray}
Here, we defined $\Delta_c = \omega_0 - \omega_p$, the coupling matrix elements $g_{\nu\mu} = g_0 \sum_i \langle \nu |2i\rangle\langle 2i | \mu \rangle$ and introduced the Fermi distribution function $n_F^j$ such that lead $j=L,R$ is described by its Fermi energy $\mu_j$ and the temperature $T$. The coupling to the environment is captured in the rates
\begin{eqnarray}
    \kappa &=& \sum_x \kappa_x \approx 2\pi \sum_x g_x^2(\omega), \\
    \tilde \Gamma_{L\mu} &=& \pi \sum_{k,\sigma,v} \left| \sum_i \langle 1i | \mu \rangle \tau_{Lk\sigma v} \right|^2 \delta(\varepsilon - \varepsilon_{kL\sigma v}), \\
    \tilde \Gamma_{R\mu} &=& \pi \sum_{k,\sigma,v} \left| \sum_i \langle 2i | \mu \rangle \tau_{Rk\sigma v} \right|^2 \delta(\varepsilon - \varepsilon_{kR\sigma v})
\end{eqnarray}
and the input fields
\begin{eqnarray}
    \tilde a_{\mathrm{in,x}} &=& \frac{-i}{\sqrt{2\pi}} \int \mathrm{d} \omega e^{-i \omega (t-t_0)} b_{x,\omega} (t_0) \\
    \tilde d_{j\mu }^\mathrm{in} (t) &\approx & \frac{-i}{\pi} \sqrt{ \frac{\tilde \Gamma_{j\mu} }{2} n_F^j(E_\mu) } \,  e^{-i (E_\mu - \Omega_\mu)(t-t_0)}.
\end{eqnarray}

The derivation of these expressions relies on the approximations that the couplings between the resonator and the environment is flat, $g_x(\omega) \approx g_x$, and that the electronic reservoir of the leads are infinite. As a reference we also define the total tunneling rates between dot $j=L,R$ and its lead,
\begin{equation}
    \Gamma_{j} \approx \pi \sum_k |\tau_{k j \sigma v} |^2 \delta(\varepsilon - \varepsilon_{j k \sigma v}),
\end{equation}
assuming that $\tau_{k j \sigma v} \approx \tau_{k j}$. For simplicity we will further choose $\Delta_c = 0$ and $t-t_0 \gg 1/\kappa,1/\tilde \Gamma_{L(R)}$ throughout the remainder of the paper.

We assume that the resonator is driven with a coherent input field from one port only, $\tilde a_{\mathrm{in},1} =\ain$, $\tilde a_{\mathrm{in},2} =0$. Unlike Ref. \cite{PhysRevB.101.155406} we proceed by formally integrating the equation for the field $\tilde a$ to
\begin{eqnarray}
    \tilde a(t) &=& \int _{t_0}^t \mathrm d t'\, e^{-\kappa (t-t')/2} \Bigg( \sqrt{\kappa _1} \ain \nonumber\\
    && -i \sum _{\mu ,\nu } g_{\mu\nu} e^{- i (\omega _p + \Omega _{\mu }-\Omega _{\nu }) t'} \tilde{d}_\mu^\dagger (t) \tilde{d}_\nu (t) \Bigg) \label{eq_field}
\end{eqnarray}
and we apply a RWA which is justified for $|\omega_0 - (E_\mu - E_\nu)| \ll E_\mu - E_\nu \approx \omega_p$ \cite{PhysRevA.98.023849}. Introducing the auxiliary variable $N_{\mu\nu}= \tilde d_\mu^\dagger \tilde d_\nu$ the remaining system of equations is Laplace transformed.

The time dependent exponential functions result in a shift in the complex frequency space \cite{DoetschLaplace} which can be expressed by defining a displacement operator $\mathcal S_x f(s) = f(s+x)$ with $\mathcal S_x^{-1} = S_{-x}$. The resulting system of linear equations can be solved for $\mathcal L \tilde d_\mu (s)$, the Laplace-transform of $\tilde d_\mu (t)$ \cite{10.2307/2690437}:

\begin{widetext}
\begin{eqnarray}
    \mathcal L \tilde d_\mu (s) &=& \sum_{\nu} (A_0^{-1})_{\mu\nu} (s) C_\nu(s) + \sum_{\nu,\lambda,\eta} (A_0^{-1})_{\mu\nu}(s) (A_1)_{\nu\lambda} (s) (A_0^{-1})_{\lambda\eta} \left[s+i(E_\lambda-E_\eta-\Omega_\lambda+\Omega_\eta)\right] C_\eta(s) \nonumber\\
    && - \sum_{\nu,\lambda,\eta}  (A_0^{-1})_{\mu\nu}(s) (A_1)_{\nu\lambda} (s) (A_0^{-1})_{\lambda\eta}^* \left[s+i(E_\lambda+E_\eta-\Omega_\lambda-\Omega_\eta)\right] C_\eta^* (s), \label{eq_LTsolution}
\end{eqnarray}
\end{widetext}
where the initial conditions at $t_0$ enter via
\begin{equation}
    C_\mu (s) = \sqrt{2\pi} \hat d_\mu^\mathrm{in}  \left[ \frac{1}{s+i(E_\mu - \Omega_\mu)} - \sum_\nu \frac{\tilde N_{\nu\nu}(t_0)}{s+2\gamma_\nu}\right] + \tilde d_\mu (t_0).
\end{equation}
and the coefficient matrices are
\begin{eqnarray}
(A_0)_{\mu\nu} (s) &=& \delta_{\mu\nu} \left\{s+ \left[i\left( E_\mu - \Omega_\mu\right) + \gamma_\mu\right]\right\} + i p_{\mu\nu},\\
(A_1)_{\mu\nu} (s) &=& 2\pi \hat d_\mu^\mathrm{in} \left( \hat d_\nu^\mathrm{in}\right)^* (s+2\gamma_v)^{-1},\\
\hat d_\mu^\mathrm{in} &=& \frac{-i}{\pi} \sum_{j=L,R} \sqrt{\frac{\tilde \Gamma_{j\mu}}{2}\, n_F^j(E_\mu)},\\
\gamma_\mu &=& \sum_{j=L,R} \left(1-n_F^j(E_\mu)\right) \tilde \Gamma_{j\mu},\\
p_{\mu\nu} &=& 4 \frac{g_0}{\kappa} |a_\mathrm{in}| \delta\left(\omega_p - \Omega_\nu + \Omega_\mu\right) + q_{\mu\nu}.\label{eq_p-mu-nu},
\end{eqnarray}
where $\delta_{\mu\nu}$ is the Kronecker symbol and $\delta$ is the $\delta$-distribution. The term $q_{\mu\nu}$ in Eq.~(\ref{eq_p-mu-nu}) can be understood as a back-action of the DQD on itself via the resonator. It can be expected to be small and is neglected in the discussion in Secs.~\ref{sec_discussion} and \ref{sec_valley}. An estimate for the case $\kappa \ll \Gamma_L,\Gamma_R$ is given by
\begin{eqnarray}
    q_{\mu\nu} & \lesssim & \sqrt{2\pi}\, g_0^2 \zeta_{\mu\nu} 2 e^{-\kappa \tau/2} \left| \sum_\lambda \hat d_\lambda^\mathrm{in} \right|^2,\\
    \zeta_{\mu\nu} &=& \mathrm{sgn}\left\{ (\Omega_\nu - \Omega_\mu) \cos \left[ (\Omega_\nu - \Omega_\mu) \tau \right] \right\}, \\
    \tau &=& 2 \mathrm{max}_{\lambda,\eta}|\Omega_\lambda - \Omega_\eta| /(\kappa \Gamma_R),
\end{eqnarray}
where $\mathrm{sgn}$ is the sign function.

Although the Laplace transform $\mathcal L \tilde d_\mu$ itself has no physical meaning, the steady state solution of the Langevin equations can be obtained from Eq.~(\ref{eq_LTsolution}) by the identity
\begin{equation}
    \lim_{t\to \infty} \tilde d_\mu(t) = \lim_{s \to 0+} s\mathcal L \tilde d_\mu(s)\label{eq_laplaceLim}
\end{equation}
if $\mathcal L \tilde d_\mu$ has no other singularities with $\mathrm{Re}\, s \geq 0$ than a simple pole at $s=0$ \cite{DoetschLaplace}. Alternatively, an inverse Laplace transform of Eq.~(\ref{eq_LTsolution}) yields the real-time dynamics of $\tilde d_\mu (t)$ for $t-t_0 \gg 1/\kappa,1/\tilde \Gamma_{L(R)}$. Substituting into Eq.~(\ref{eq_field}) gives the time evolution of $\tilde a$, the output field $\aout$ is obtained from the IO relations \cite{PhysRevA.30.1386}. Eventually, the current from the DQD to the drain contact in units of electron charges $e$ is given by $I/e = \frac{\mathrm{d}}{\mathrm{d}t}  \sum_{k,\sigma,v} \left\langle c_{kR\sigma v}^\dagger c_{kR\sigma v} \right\rangle$, which can be expressed using the IO relation \cite{PhysRevB.101.155406},
\begin{equation}
    I = 2 e\sum_{i,\mu} |\langle 2i | \mu \rangle|^2 \left( \Gamma_{R} \left\langle \tilde d_\mu^\dagger \tilde d_\mu \right\rangle - \sqrt{2\pi} \mathrm{Re}  \left\langle \tilde d_\mu^\dagger \tilde d_{R\mu}^\mathrm{in} \right\rangle \right). \label{eq_current}
\end{equation}
The first term in Eq.~(\ref{eq_current}) describes the decay of population in the right QD to the right lead with a decay rate $\Gamma_R$ computed from IO theory. The second term describes tunneling from the right lead to the DQD which becomes relevant with high temperatures or a small bias window $\mu_L - \mu_R$. We emphasize that the solution using Eq.~(\ref{eq_laplaceLim}) is analytic, although it requires the (numerical) diagonalization of $H_\mathrm{QD}$.

It is also possible to solve the Laplace-transformed equations without RWA. The analogue of Eq.~(\ref{eq_LTsolution}) beyond the RWA is presented in Appendix~\ref{app_beyRWA}. We further note that it is straightforward to generalize our theory to a QD system with arbitrary geometry coupled to any number of resonators.

\section{Discussion\label{sec_discussion}}

We first investigate the basic properties of the interaction without valley degree of freedom. In this case the (1,1) triplets ($T_{0(\pm)}$) are blockaded unless spin-flip processes allow transitions to the (0,2) singlet ($S^{(0,2)}$). The solution with RWA is used and the stationary state is obtained from Eq.~(\ref{eq_laplaceLim}). We assume a bias window of $\mu_L - \mu_R = \SI{1}{\milli\electronvolt}$, centered around the DQD levels and $T = \SI{0.1}{\kelvin}$.

\begin{figure}
\begin{center}
\includegraphics[width=0.5\textwidth]{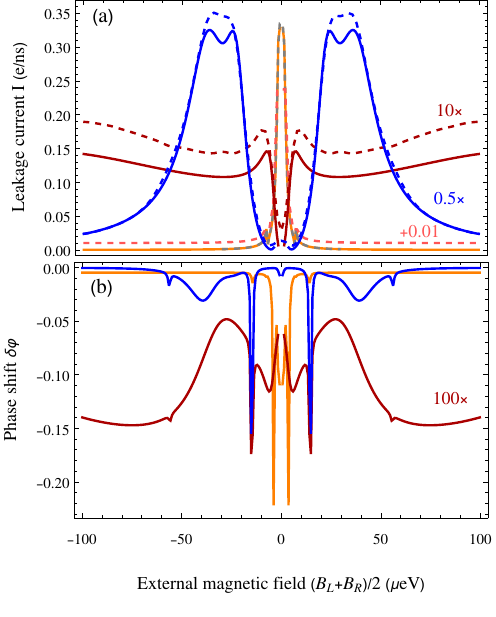}
\caption{Comparison of the leakage current and the cavity response in different regimes. (a) Leakage current $I$ as a function of the Zeeman splitting. The orange curve has small tunneling $t_c^2+t_f^2 = \SI{1}{\micro\electronvolt}^2$, $B_R-B_L = \SI{14}{\micro\electronvolt}$ and $\omega_0 = \sqrt 2\, \SI{5}{\micro\electronvolt}$. The red (blue) curve have strong tunneling $t_c = \SI{20}{\micro\electronvolt}$ and the magnetic field gradient $B_R-B_L = \SI{1.5(10)}{\micro\electronvolt}$ and $\omega_0 = \sqrt 2\, \SI{20}{\micro\electronvolt}$.
For all curves we chose $t_c = 4t_f$, $\eps = U$, $g_0 = \SI{0.5}{\micro \electronvolt}$, $\kappa = \SI{1.28}{\micro\electronvolt}$, $\Gamma_L = 2\Gamma_R= \SI{10}{\micro\electronvolt}$, $\ain = \SI{1}{\micro\electronvolt^{1/2}}$. 
The dashed curves have the same setting as their solid counterparts (light red corresponds to orange) except $\kappa = \SI{0.0128}{\micro\electronvolt}$ to highlight the back-action of the photons on the DQD.
The dashed gray curve has the same setting as the orange curve but is obtained from a numerical treatment of the system, the agreement with the analytically approximated solution is very good. 
For clarity the red (blue) curves are scaled up (down) by a factor of 10 (0.5) and the light red curve is offset by $0.01e/\si{\nano\second}$. Note that $\Gamma_R$ is relatively small, here.
(b) Phase shift $\delta\varphi$ of outgoing photons in the same cases. Up to narrow resonances $E_\mu - E_\nu = \omega_p$ the phase shift correlates with the current $I$; the connection is established by Eqs.~(\ref{eq_field}) and (\ref{eq_current}).\label{fig_novalley}}
\end{center}
\end{figure}

In Fig.~\ref{fig_novalley}(a) the leakage current $I$ is plotted for different values of the total tunneling strength and the differences in Zeeman splitting, $B_L-B_R$. The dashed gray curve is based on an exact numerical treatment without RWA of the system for the same parameters as the orange curve. Analogously, in the case with valley pseudospin a comparison between the analytical solution and the exact numerical treatment without RWA is shown in Fig.~\ref{fig_valleyChar}(a). The agreement with the approximate analytical solution is very good in both cases. However, in the case of Fig.~\ref{fig_valleyChar}(a) the deviation is larger since the RWA is worse since $\varphi_v$ is relatively close to $\pi/4$. Note that the analytically derived curves are missing one point at $B_L+B_R=0$. These results are obtained from Eq.~(\ref{eq_laplaceLim}), however, this identity is only valid if $\mathcal{L}\tilde d_\mu(s)$ has no poles with $\mathrm{Re}\,s\geq 0$ other than a simple pole at $s=0$~\cite{DoetschLaplace}. Due to the magnetic field dependence of the triplet energies at $B_L+B_R=0$ a pole with multiplicity two also falls to $s=0$ and therefore Eq.~(\ref{eq_laplaceLim}) must not be applied. We emphasize that an inverse Laplace transform of Eq.~(\ref{eq_LTsolution}) can still yield the solution at these instances, although at a higher computational cost.

As Fig.~\ref{fig_novalley}(a) shows, the model can reproduce the features known to appear in spin blockade. The leakage current exhibits a transition from a Lorentzian dip to a peak at zero magnetic field when the tunneling is decreased \cite{PhysRevB.80.041301}. For large magnetic fields the current decays to zero \cite{PhysRevB.84.245407,Lai-SciRep-2011} to the extent that it can appear as a double peak with the width determined by $\Gamma_R $ \cite{PhysRevB.95.155416}. We also observe a small dip at $B_L+B_R=\pm 2\sqrt{2}t_c$ where the strongly hybridized singlets anticross with the $T_\pm$ triplet states that are only weakly coupled to the $(0,2)$ sector since the degenerate levels rearrange into blockaded and open states \cite{PhysRevB.80.041301}. 

The phase shift $\delta\varphi = \arg\; \aout / \ain$, displayed in Fig.~\ref{fig_novalley}(b), responds to the current since the photons couple to the dipole moment of the tunneling electrons, as can be expected from Eqs.~(\ref{eq_field}) and (\ref{eq_current}). Thus, it is possible to qualitatively infer the relative change of the leakage current from $\delta\varphi$ during the sweep. However, to estimate the value of $I$ it is important to respect the dependence of $g_{\mu\nu}$ of the levels $\mu,\nu$ that carry current on the DQD parameters, and to consider how close $\omega_p$ is to a resonance with the DQD level splitting. In the example this is highlighted by comparing the cases of strong (blue) and weak (orange) tunneling both with strong magnetic gradient. Even though the maximal current in the blue curve is twice as high as the maximum of the orange curve, the resonator response is weaker due to different effective couplings.

The resonator response $\delta\varphi$ also exhibits narrow peaks where the probe field is resonant with a DQD level transition, $\omega_p = E_\mu - E_\nu$. These resonances could disturb a resonator-aided measurement of the leakage current if the level structure of the DQD is unknown.

Furthermore, there can be a significant back-action of the resonator photons on the current $I$. The dashed curves in Fig.~\ref{fig_novalley}(a) strongly deviate from their solid counterparts, although the only difference is a smaller value of $\kappa$. The reason is  that absorption or emission of photons can lead to a different electronic equilibrium state than without the resonator. In particular, near the avoided level crossings (ALCs) between singlet and triplet states the electron can be excited from a triplet to a singlet which has much higher probability for a transition to the drain. Vice versa, excitation from a singlet to a triplet reduces the current.

To estimate the magnitude of this effect we treat the ALCs between the singlets and the $T_0$ ($T_\pm$) at $B_L+B_R= 0$ ($B_L+B_R=\pm 2\sqrt{2}t_c$) separately with an effective three (two) level Hamiltonian. In both cases we find that the current has the form
\begin{equation}
    I' \approx I_0 + p_0 \sqrt{I_0}\delta\label{eq_backaction}
\end{equation}
with $p_0 = 4 g_0 |\ain|/\kappa$, the current $I_0$ through the uncoupled DQD and a correction $\delta$. Thus, in strongly driven high-Q resonators it can be expected that the current is altered by the resonator. The explicit expressions that describe the relative change of $I$ near these ALCs are given in Appendix~\ref{app_backaction}.

We further find that in the off-resonant dispersive regime $g_{\mu\nu} \ll |\omega_p - (E_\mu - E_\nu)|$ \cite{PhysRevA.98.023849} the leakage current $I(\eps)$ appears to be displaced along the detuning axis $\eps$ if $p_0$ is large. This can be understood by the well-known dispersive shift of the energy splittings to which the DQD-photon interaction is reduced in the dispersive regime. The shift $\chi_m (a^\dagger a + 1/2)$ of the molecular transition frequencies \cite{PhysRevB.100.245427} affects only the (0,2) singlet state and appears as a shift of $\eps \to \eps - 2 \chi_m (a^\dagger a + 1/2)$ near the ALC of the singlets at $\eps = 0$.

\section{Lifted Valley Degeneracy\label{sec_valley}}

In this section the repercussions of the valley degree of freedom are discussed. This case is important for conduction-band electron transport through silicon DQDs with near-degenerate valleys. Eventually, we present schemes to estimate the leakage current in Sec.~\ref{sec_valleyObs} and to measure the parameters of the (valley) Hamiltonian from the resonator response during a transport experiment in Sec.~\ref{sec_valleyChar}.

Now, there are 16 (1,1) states with subtle conditions to be blockaded \cite{PhysRevB.80.201404,PhysRevB.82.155424,PhysRevB.82.155312,Hao2014}. The effects discussed in Sec.~\ref{sec_discussion} -- a resonator response to the leakage current and back-action of both resonant and off-resonant photons in the current for large $p_0$ -- are present in this case as well. However, in the more complex level diagram there are many transitions between different supersinglet and -triplet states that can interact with the resonator.

To identify the resonance conditions we use the analytical result Eq.~(\ref{eq_laplaceLim}) and determine the contribution of each eigenstate to the leakage current and the transmission. We approximate the relevant eigenstates of $H_\mathrm{QD}$ by transforming the total Hamiltonian $H$ into a rotating frame describing resonant transitions between these states, performing a RWA and diagonalizing the part of the Hamiltonian acting on the DQD. Further assuming $t_c \gg t_f$ we find that any of the following resonance conditions can give rise to a strong resonator response.
\begin{eqnarray}
    \omega_p &\approx & \sqrt{2} |t_c \cos\varphi_v|, \label{eq_resonanceCrudeApprox}\\
    \omega_p^2 &\approx & \left[(\varepsilon-U) \pm \frac{B_L-B_R}{2} \pm \frac{\Delta_L+\Delta_R}{2} \right]^2 \nonumber \\
    && +4(t_c \sin\varphi_v)^2,  \label{eq_firstOtherResonance} \\
    \omega_p^2 &\approx & \left[(\varepsilon-U) \pm \frac{B_L+B_R}{2} \pm \frac{\Delta_L+\xi\Delta_R}{2} \right]^2 \nonumber \\
    && + 4(t_f \lambda_\xi)^2 \label{eq_middleOtherResonance} 
\end{eqnarray}
with $\xi = \pm 1$, $\lambda_{-1} = \cos\varphi_v$, $\lambda_1 = \frac{1}{\sqrt{2}}\sin\varphi_v$ and also
\begin{equation}
    \omega_p^2 \approx \left[(\varepsilon-U)-\xi \Delta_R \pm \frac{B_L-B_R}{2} \right]^2 +4(t_f\lambda_\xi)^2 \label{eq_lastResonance}
\end{equation}
with $\xi = 0,\pm 1$, $\lambda_{\pm 1} = \frac{1}{\sqrt{2}}\sin\varphi_v$, $\lambda_0 = \cos\varphi_v$. Note that Eq.~(\ref{eq_resonanceCrudeApprox}) is a good approximation only for $\eps = U$. A more accurate expression is given in Appendix~\ref{app_resonanceCondition}. 

\begin{figure}
\begin{center}
\includegraphics[width=0.5\textwidth]{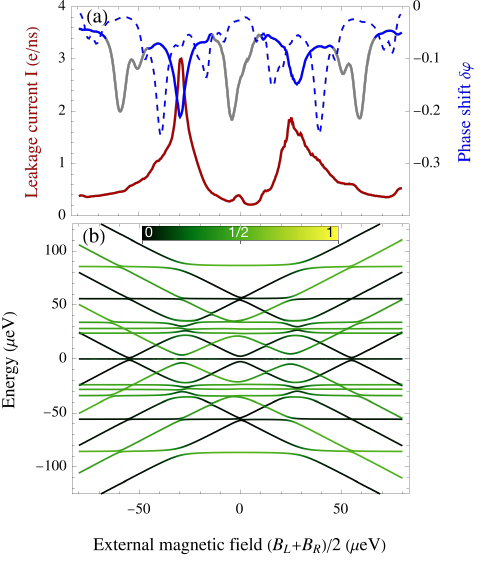}
\caption{The phase shift $\delta\varphi$ can be used to estimate the leakage current, except in a window set by the resonator linewidth $\kappa$ around the ALCs between supertriplets with opposite spin, given in Eqs.~(\ref{eq_valley_meas_ALC1})-(\ref{eq_valley_meas_ALC3}). (a) Current $I$ (red) and phase shift $\delta\varphi$ (blue, gray) with valley. The solid (dashed) curve of $\delta\varphi$ is computed with $\omega_p = \SI{60}{\mu\electronvolt}$ ($\omega_p = \SI{40}{\mu\electronvolt}$) and thus the resonances Eq.~(\ref{eq_resonanceCrudeApprox})-(\ref{eq_lastResonance}) are met at different values of $B_L+B_R$. The choice for the solid curve is according to the scheme to measure $I$ discussed in Sec.~\ref{sec_valleyObs}, the intervals with gray curve are those where the measurement cannot yield clear information on $I$.
(b) Level diagram of the DQD. The color scale indicates the overlap with the (0,2) sector, $\sum_i |\langle \mu | 2i \rangle |^2$: green (black) states are open (blockaded).
Parameters for both panels are $\Delta_L = \SI{50}{\micro\electronvolt}$, $\Delta_R = \SI{60}{\micro\electronvolt}$, $\varphi_v = \pi/13$, $t_c = 4 t_f =\SI{20}{\micro\electronvolt}$, $B_L - B_R = 2\kappa = \SI{10}{\micro\electronvolt}$.
The rest is as in Fig.~\ref{fig_novalley}.\label{fig_valleyObs}}
\end{center}
\end{figure}

Due to the different couplings $g_{\mu\nu}$ the resonances Eq.~(\ref{eq_resonanceCrudeApprox}-\ref{eq_lastResonance}) have different visibilities in the resonator response. This can be seen in Fig.~\ref{fig_valleyObs}(a) where $I$ is plotted together with $\delta\varphi$ for two different values of $\omega_p$.

\subsection{Observation of the Leakage Current\label{sec_valleyObs}}

During a sweep of $\eps$, the Zeeman splitting or the tunneling, several of the resonances Eqs.~(\ref{eq_resonanceCrudeApprox})-(\ref{eq_lastResonance}) can be traversed. As a result, similar values of $I$ can appear with different visibility in $\delta\varphi$ since the relevant resonances are associated with different dipole moments. In the example of Fig.~\ref{fig_valleyObs}(a), the dashed curve is not suited to extract information on $I$ since the resonances are met on the flanks of the peaks of $I$.

This challenge can be partially mitigated if the measurement is performed with $\eps \approx 0$, $\kappa \gtrsim |B_L - B_R|/2$ and $\omega_p \approx (\Delta_L + \Delta_R)/2 + t_f$. With this choice of $\omega_p$ the probe field is approximately resonant with the splitting between pairs of states where both states have equal spin projection but opposite valley configuration. The eigenenergies of such pairs are separated by the valley splitting and are parallel as a function of $B_L+B_R$ (e.g. Fig.~\ref{fig_valleyObs}(b)). The resonator field is therefore sensitive to ALCs which lift the blockade if one state from such a pair is involved. The choice of $\kappa$ makes sure that all states are part of one such pair. This is necessary since the supertriplet states without spin polarization are shifted by the difference in Zeeman splitting. An advantage of a relatively large $\kappa$ is the suppression of unwanted back-action effects.

The result of a sweep of $B_L + B_R$ is depicted in Fig.~\ref{fig_valleyObs}(a) by the solid curve. The figure also highlights the limitation of this method. Near the ALCs of supertriplets with opposite spin the correlation of resonator response and leakage current is broken and the phase shift is much stronger as can be expected from $I$. These ALCs occur at
\begin{eqnarray}
    B_L+B_R &=& 0,\label{eq_valley_meas_ALC1}\\
    B_L+B_R &=& \frac{B_L-B_R}{2}, \\
    B_L+B_R &=& \pm (\Delta_L+\Delta_R + 2 t_f).\label{eq_valley_meas_ALC3}
\end{eqnarray}
Within a window of width $\approx 4 \kappa$ around these ALCs the resonator-aided measurement of the leakage current is not reliable. These intervals are indicated in Fig.~\ref{fig_valleyObs}(a) by changing the color of the curve to gray.

Thus, the utility of the proposed measurement scheme to observe the leakage current is limited by the mean valley splitting $(\Delta_L + \Delta_R)/2$ and the difference in Zeeman splitting $|B_L - B_R|$ to which $\kappa$ is tied. Another practical limitation to certain regimes might arise from the requirement to set $\omega_p$ to a value determined by the valley splitting. In silicon-based heterostructures the valley splitting is sensitive to the fabrication details  \cite{PhysRevB.80.081305,PhysRevB.84.155320,PhysRevResearch.2.043180} and electrically tunable only in a limited range \cite{PhysRevResearch.2.043180,PhysRevApplied.13.034068,Yang2013,doi:10.1063/1.4972514}, in bilayer graphene the valley splitting can be tuned by means of an out-of-plane magnetic field \cite{doi:10.1021/acs.nanolett.0c04343} which also couples to the spin magnetic moment. Prior knowledge of the valley splittings required to identify the operating regime of this measurement technique can be obtained from a transmission measurement in a closed system~\cite{PhysRevB.94.195305,PhysRevLett.119.176803,Russ_2020,Borjans2021}.

\subsection{DQD Characterization\label{sec_valleyChar}}

\begin{figure}
\begin{center}
\includegraphics[width=0.5\textwidth]{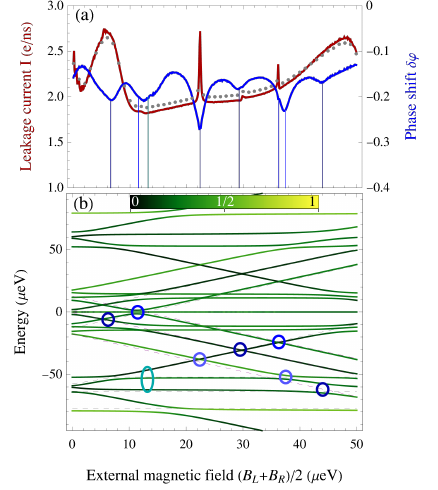}
\caption{Outline of the proposed measurement of unknown DQD parameters. (a) The phase shift $\delta\varphi$ for $\omega_p = \sqrt{2} |t_c \cos\varphi_v|$ as a function of the magnetic field. Note that here $\omega_p$ is not at the ideal point for measuring $I$ (Sec.~\ref{sec_valleyObs}), thus $\delta\varphi$ cannot be used to reliably quantify $I$. The gray dotted line shows the exact numerical solution for $I$. The agreement with the analytical solution is very good, although not as good as in Fig.~\ref{fig_novalley}, since $\varphi_v$ is close to $\pi/4$ (large inter-valley tunneling). (b) The level diagram of the DQD as a function of the magnetic field. As indicated by the blue vertical lines and matching ellipses the extrema of $\delta\varphi$ can be related to the ALCs between $\tilde E_1 - \tilde E_3$ (dashed purple) and $\tilde E_1' - \tilde E_6'$ (dashed gray) given in Appendix~\ref{app_characterization}. Additional resonances (cyan ellipse) or very close ALCs can obscure the result. A useful technique to identify the undesired resonances is sketched in Fig.~\ref{fig_resonancePositions}.
Parameters for both panels are the same as Fig.~\ref{fig_valleyObs} up to $\omega_p = \sqrt 2\, \SI{10}{\micro\electronvolt}$ and $\varphi_v = \pi/3$. The color scale indicates the overlap with the (0,2) sector, $\sum_i |\langle \mu | 2i \rangle |^2$. Due to the strong valley-flip tunneling none of the states is totally blockaded (black).\label{fig_valleyChar}}
\end{center}
\end{figure}

It is favourable to know the valley splittings $\Delta_{L(R)}$ and the valley phase $\varphi_v$ for a given DQD device. In the context of Pauli blockade this is important since the ratio of the different tunneling matrix elements has a major effect on $I$, as a comparison of the leakage current in Figs.~\ref{fig_valleyObs} and \ref{fig_valleyChar} shows. Furthermore, when using the scheme discussed in Sec.~\ref{sec_valleyObs} to measure the leakage current knowledge about the valley splitting is crucial to set $\omega_p$ and to determine the windows of unreliable results that should be clipped. 

This knowledge can be inferred from a prior measurement using well-established protocols and the same microwave resonator \cite{PhysRevB.94.195305,PhysRevLett.119.176803,Russ_2020,Borjans2021}. Here, we present an alternative resonator-aided scheme for the DQD characterization during a transport experiment.

For this application the resonance condition Eq.~(\ref{eq_resonanceCrudeApprox}) can be used, $\omega_p \approx \sqrt{2} |t_c \cos\varphi_v|$, with $\eps = U$. Thus, the probe field is resonant with the splitting between the energy levels $\hat E_{1(3)}$ and $ \hat E_2$ approximately given by Eq.~(\ref{eq_E1}-\ref{eq_E2}) in Appendix~\ref{app_characterization}. The ALCs of these states with $\hat E_1' - \hat E_6'$, Eq.~(\ref{eq_E1p}-\ref{eq_E6p}), during a sweep of the magnetic field give rise to an extremum in the resonator response, each. This is depicted in Fig.~\ref{fig_valleyChar}. Unknown DQD parameters can then be inferred by equating the expressions from Appendix~\ref{app_characterization} and solving for the unknown DQD parameters.

The performance of this scheme is limited by two possible caveats. First, the ALCs might be closer than the resonator linewidth $\kappa/2$ and could thus not be resolved individually. This is shown in Fig.~\ref{fig_valleyChar} near $(B_L + B_R)/2 = \SI{37}{\micro\electronvolt}$. Second, some of the other resonance conditions, Eqs.~(\ref{eq_firstOtherResonance})-(\ref{eq_lastResonance}), might be met during the sweep, giving rise to an unexpected extremum. This is shown by the cyan highlight in Fig.~\ref{fig_valleyChar}. 

To identify the expected extrema of $\delta\varphi$ and distinguish them from undesired features it is of avail to sweep the tunnel coupling $t_c$ in a small range. The measurement scheme relies on the resonance condition $\omega_p \approx \sqrt{2} |t_c \cos\varphi_v|$, thus, if $t_c$ is swept the visibility of the desired lines changes with a Lorentzian profile centered around the resonance with a width set by the resonator linewidth. Furthermore, the position of the ALCs depends on $t_c$, thus the positions of the extrema along the magnetic field axis is a function of $t_c$, this can aid the discrimination of very close extrema. Among the undesired resonance conditions only Eq.~(\ref{eq_firstOtherResonance}) depends on $t_c$. Features in $\delta\varphi$ caused by Eq.~(\ref{eq_firstOtherResonance}) appear at different values of $B_L+B_R$ if $t_c$ is changed, however, with approximately constant visibility. The other undesired resonances, Eqs.~(\ref{eq_middleOtherResonance}-\ref{eq_lastResonance}) do not depend on $t_c$, thus they remain at the same magnetic field value and approximately maintain their visibility during the sweep of $t_c$. The different behaviour of desired and undesired extrema makes it possible to identify them. For the example of Fig.~\ref{fig_valleyChar} this is illustrated in Fig.~\ref{fig_resonancePositions}.

\section{Summary and Conclusion\label{sec_summary}}

In this article we extended the generalized IO theory and derived an analytic description of electronic transport in the Pauli blockade regime in semiconductor quantum dots coupled to a microwave resonator. We first investigated the interaction of a spin blockade with the microwave photons within a RWA, although we also provide a solution beyond the RWA. While the resonator's output field carries quantitative information on the leakage current, there can also be back action on the current. Near the resonance of the probe field with a DQD transition, this is due to the absorption of photons. Away from resonance the mutual dispersive shift of DQD and resonator may also obscure experimental results. We analytically estimated the change of the leakage current and concluded that back-action can be mitigated by choosing parameters where $p_0 = 4 g_0 |\ain|/\kappa$ is small.

In the case of a lifted valley degeneracy, i.e. for silicon or carbon based spin qubits, the back-action effects persist. The resonator response to the leakage current, however, can show a complicated dependence due to different resonance conditions with a large number of states. As a result, there is not necessarily a quantitative agreement between $\delta\varphi$ and $I$. Nonetheless, we devised a scheme that allows to observe the leakage current from a measurement of the output field, limited by the valley splitting and the difference in Zeeman splitting between the QDs. Furthermore, we provide a scheme that can be used to extract information on unknown DQD parameters simultaneous to a transport experiment.

Pauli blockade is a powerful tool for the characterization of spin qubits. Our results can help leveraging its utility to large-scale qubit applications without dedicated components for charge or current sensing in each QD. This can be useful because the same resonator used for two-qubit gates and possibly readout can accomplish this task. The back-action described here can open a novel pathway to manipulate the electronic state of a QD system by enhancing or suppressing a leakage current. For future research directions, applications relevant for photonic platforms are also worth investigating. For example, a tunable interaction between two resonators coupled to the same DQD is conceivable, harnessing the properties of Pauli blockade and the back-action.


\begin{acknowledgments}
We thank Jonas Mielke, Benedikt Tissot, Philipp Mutter, Monika Benito and Jeroen Danon for helpful discussions. Florian Ginzel acknowledges a scholarship from the Stiftung der Deutschen Wirtschaft (sdw) which made this work possible. This work has been supported by the Army Research Office (ARO) grant number W911NF-15-1-0149.
\end{acknowledgments}


\appendix
\section{Solution beyond the RWA\label{app_beyRWA}}

The solution of the Langevin equations derived in the main text and used for the discussion of the interaction was based on a rotating wave approximation (RWA). The RWA is not strictly necessary to solve the problem. Here, we give the analogue of Eq.~(\ref{eq_LTsolution}) without RWA:
\begin{widetext}
\begin{eqnarray}
    \mathcal L \tilde d_\mu (s) &=& \sum_\nu (\hat A_0^{-1})_{\mu\nu}(s) C_\nu(s) - \frac{2i}{p_0} \sum_\nu \left\{ C_\nu[s- i(\omega_p-\Omega_\nu-\Omega_\mu)] + C_\nu[s+ i(\omega_p-\Omega_\nu-\Omega_\mu)] \right\}\\
    && + \sum_\lambda \Big\{ \sum_\nu (\hat A_0^{-1})_{\mu\lambda}(s) (A_1)_{\nu\lambda}(s) - \frac{2i}{p_0} \sum_\nu \left\{ (A_1)_{\nu\lambda}[s- i(\omega_p-\Omega_\nu-\Omega_\mu)] + (A_1)_{\nu\lambda}[s+ i(\omega_p-\Omega_\nu-\Omega_\mu)] \right\} \Big\} \nonumber\\
    && \sum_\eta \Bigg\{
    (\hat A_0^{-1})_{\lambda\eta}[s+i(E_\lambda - E_\eta - \Omega_\lambda + \Omega_\eta)] C_\eta(s) -(\hat A_0^{-1})_{\lambda\eta}^*[s+i(E_\lambda + E_\eta - \Omega_\lambda - \Omega_\eta)] C_\eta(s)^\dagger \nonumber\\
    && -\frac{2i}{p_0} \sum_\zeta \big\{ C_\zeta[s+i(E_\lambda - E_\eta - \Omega_\lambda + \Omega_\zeta - \omega_p)] + C_\zeta[s+i(E_\lambda + E_\eta - \Omega_\lambda + 2\Omega_\eta - \Omega_\zeta + \omega_p)] \nonumber\\
    && + C_\zeta^\dagger[s+i(E_\lambda + E_\eta - \Omega_\lambda - 2\Omega_\eta + \Omega_\zeta - \omega_p)] + C_\zeta\dagger[s+i(E_\lambda + E_\eta - \Omega_\lambda - \Omega_\zeta + \omega_p)] \big\} \nonumber
    \Bigg\},
\end{eqnarray}
\end{widetext}
with the new definitions
\begin{eqnarray}
    (\hat A_0)_{\mu\nu} &=& \delta_{\mu\nu} \left\{ s+ \left[ i \left( E_\mu - \Omega_\mu \right) + \gamma_\mu \right] \right\}, \\
    p_0 &=& 4 g_0 |\ain|/\kappa.
\end{eqnarray}

\section{Estimation of the back-action\label{app_backaction}}

To assess the magnitude of the back-action if $\omega_p$ is resonant to a singlet ($S$)-triplet ($T$) ALC we use reduced Hamiltonians of the states contributing to these ALCs. In both cases Eq.~(\ref{eq_laplaceLim}) yields the current
\begin{eqnarray}
    I' &\approx & \frac{2\Gamma_R}{\hbar} \frac{|k_c|^2}{N^2}, \label{eq_backaction_explicit}\\
     k_c &=& \left( \beta + \frac{p_0}{\alpha} \beta' \right) \prod_\mu N_\mu^{-1}
\end{eqnarray}
with the following definition:
\begin{eqnarray}
    \alpha &=& \sum_\mu |\hat d_\mu^\mathrm{in}|^2 \prod_{\nu \neq \mu}  \gamma_\nu (E_1 - i \gamma_\nu) \nonumber \\
    && + \prod_\mu \frac{\gamma_\mu}{\pi} (E_1 - i \gamma_\mu)\\
    \beta &=& \sum_\mu \hat d_\mu^\mathrm{in} \prod_{\nu \neq \mu} N_\nu (E_1 - i \gamma_\nu)\\
    N^2 &=& \sum_\mu \pi |\hat d_\mu^\mathrm{in}|^2 \prod_{\nu \neq \mu} \gamma_\nu (E_1^2 + \gamma_\nu^2).
\end{eqnarray}

Near $B_L+B_R=0$ with the three-level system
\begin{equation}
    H_{T_0} =
    \begin{pmatrix}
    0 & \frac{B_L - B_R}{2} & 0 \\
    \frac{B_L - B_R}{2} & 0 & \sqrt2 t_c^* \\
    0 & \sqrt 2 t_c & U-\eps
    \end{pmatrix}\label{eq_HamiltonianST-1}
\end{equation}
in the basis $\{ |T_0^{(1,1)}\rangle, |S^{(1,1)}\rangle, |S^{(0,2)}\rangle \}$ it is furthermore
\begin{eqnarray}
    \beta' &=& \alpha \left[ N_1 \hat d_1^\mathrm{in} \sum_{\mu = 1,2} N_\nu (E_1 - i \gamma_\nu) + N_2 N_3 \delta \right] \nonumber\\
    && + \beta \Big[ \gamma_2 \gamma_3 (\hat d_1^\mathrm{in})^* \delta - i \gamma_1 \hat d_1^\mathrm{in} (\gamma_3 (\hat d_2^\mathrm{in})^* (i E_1 \gamma_3) \nonumber\\
    && + \gamma_2 (\hat d_3^\mathrm{in})^* (i E_1 \gamma_2)) \Big]\\
    \delta &=& \hat d_2^\mathrm{in}(E_1 - i \gamma_3) + \hat d_3^\mathrm{in}(E_1 - i \gamma_2) \\
    N_\mu &=& \Bigg[ 4\Bigg| \frac{t_c}{B_L-B_R +E_1} \Bigg|^2 + \Bigg| \frac{E_1+ (\mu -1) \omega_p + 2 (\eps - U)}{2 t_c} \nonumber\\
    && - \frac{2 t_c^*}{B_L-B_R + E_1 + (\mu-1)\omega_p} \Bigg|^2 +1 \Bigg]^{1/2}.\\
\end{eqnarray}
Expanding this result to first order in $p_0$ results in the form of Eq.~(\ref{eq_backaction}).

Analogously, near the ALCs of the $T_\pm$ triplets with each singlet branch approximated by the two-level system
\begin{equation}
    H_{T_\pm} =
    \begin{pmatrix}
    \pm \frac{B_L + B_R}{2} & t_f^* \\
    t_f & U-\eps
    \end{pmatrix}\label{eq_HamiltonianST-2}
\end{equation}
in the basis $\{ |T_\pm^{(1,1)}\rangle, |S^{(0,2)}\rangle \}$ it is
\begin{eqnarray}
    \beta' &=&  \left(\alpha - \prod_\mu \frac{\gamma_\mu}{\pi} (E_1 - i \gamma_\mu) \right) \left( \gamma_1 N_2 (\hat d_2^\mathrm{in})^* - \gamma_2 N_1 (\hat d_1^\mathrm{in})^* \right) \nonumber\\
    && - \frac{i}{\pi} \sum_\mu N_\mu \hat d_\mu^\mathrm{in} \prod_\mu \gamma_\mu (E_1 - i \gamma_\mu)\\
    N_\mu &=& \frac{1}{4}\sqrt{\left| E_1 +(\mu-1)\omega_p + \eps -U \right|^2 / |t_f|^2 +1 }.
\end{eqnarray}
$E_1$ is the lowest energy eigenstate of the respective Hamiltonian.

Note that this is not the absolute leakage current, since the other states were disregarded. Nonetheless, Eq.~(\ref{eq_backaction_explicit}) can be used to estimate the relative change of $I$.

\section{More accurate expression for Eq.~(\ref{eq_resonanceCrudeApprox})\label{app_resonanceCondition}}

The resonance condition in Eq.~(\ref{eq_resonanceCrudeApprox}) is valid only for $\eps \approx U$. If the DQD energy levels are detuned the resonance condition is given by $\omega_p \approx \lambda_3 - \lambda_2$. Here, $\lambda_4 \geq \lambda_3 \geq \lambda_2 \geq \lambda_1$ are the roots of the polynomial
\begin{eqnarray}
    && s_\Delta 2 \alpha^2 \big[(\Delta_L - \Delta_R)^2 - z\big] |t_c|^2 \sin \varphi_v + (z + |\Delta_L + \Delta_R|) \nonumber\\
    && \Big\{ 
    s_\Delta (\Delta_L - \Delta_R)^2 |t_c|^4 \cos^2 \varphi_v \sin^2 \varphi_v [2\alpha - |t_c|^2 (\Delta_L \nonumber\\
    && + \Delta_R + s_\Delta z) \sin^2 \varphi_v]
    - z [|t_c]^2 \cos^2 \varphi_v (-2\alpha (\Delta_L + \Delta_R) \nonumber\\
    && |t_c|^2 \sin^2 \varphi_v)^2 - \alpha z (\alpha (B_R - \eps - U + \lambda)+ 2 s_\Delta \Delta_L \Delta_R \nonumber\\
    && |t_c]^2 \sin^2 \varphi_v) ]
    +(\Delta_L - \Delta_R)^2 [-\alpha^2 (B_R + \eps - U + \lambda) \nonumber\\
    && + s_\Delta |t_c|^2 \sin^2 \varphi_v (-2 \alpha \Delta_L \Delta_R + \cos^2 \varphi_v (2 \alpha |t_c|^2 \nonumber\\
    && - (\Delta_L + \Delta_R)|t_c|^4 \sin^2 \varphi_v))]
    \Big\} =0,
\end{eqnarray}
where we defined
\begin{eqnarray}
    \alpha &=& -2 (\Delta_L + \Delta_R) |t_c|^2 \cos \varphi_v + \Delta_L \Delta_R [\Delta_L + \Delta_R \nonumber \\
    && + s_\Delta (B_L - B_R +2 \eps -2 U)] \\
    z &=& B_L + B_R + 2 \lambda \\
    s_\Delta &=& \mathrm{sgn}(\Delta_L+\Delta_R)
\end{eqnarray}

\section{Details of the DQD characterization\label{app_characterization}}

To determine parameters of the DQD from the resonator response as described in Sec.~\ref{sec_valleyChar} the resonator is tuned to a specific resonance to be sensitive to the ALCs of the energy levels given in this section. The expressions are approximate solutions, neglecting the matrix elements that open the ALCs between them and assuming $|\Delta_L - \Delta_R|^2 \ll t_c^2 + t_f^2 \ll \Delta_L^2, \Delta_R^2$

\begin{eqnarray}
    \hat E_{1(3)} &\approx & -\frac{1}{4}\Bigg( B_L  - (-1)^\alpha \sqrt{(B_L-B_R)^2 + 32 t_c^2 \cos^2 \varphi_v} \nonumber\\
    && + 3B_R  - \frac{ (-1)^\alpha 2 (\Delta_L-\Delta_R)^2}{\sqrt{(B_L-B_R)^2 + 32 t_c^2\cos^2\varphi_v}}\Bigg) \label{eq_E1}\\
    \hat E_2 &\approx & -\frac{B_L+B_R}{2} - \frac{(\Delta_L-\Delta_R)^2}{\sqrt{(B_L-B_R)^2 + 32 t_c^2 \cos^2 \varphi_v}}, \label{eq_E2}
\end{eqnarray}
where $\alpha =0,1$. Due to the resonance condition of the measurement procedure it is $\hat E_{1(3)} \approx \hat E_2 \pm \omega_p$. The equations to determine unknown DQD parameters are obtained by equating the previous expressions with
\begin{eqnarray}
    \hat E_1' &\approx & - \frac{1}{4}\Bigg( \sqrt{(\Delta_L-\Delta_R)^2 + 32 t_c^2 \cos^2 \varphi_v} \nonumber \\ 
    && + \Delta_L + 3 \Delta_R - \frac{2t_c^2\sin^2\varphi_v}{\Delta_L+\Delta_R} \Bigg), \label{eq_E1p}\\
    \hat E_{2(3)}' &\approx & - \frac{\Delta_L + \Delta_R}{2} - \frac{t_c^2\sin^2\varphi_v}{\Delta_L + \Delta_R} \nonumber\\
    && \pm \frac{\sqrt{(B_L-B_R)^2(\Delta_L + \Delta_R)^2 + 4 t_c^2 \sin^2 \varphi_v}}{2(\Delta_L + \Delta_R)}, \\
    \hat E_4' &\approx &  \frac{1}{4}\Bigg\{ B_L + B_R - \Delta_L - 3\Delta_R + \Bigg[16 |t_f|^2 \cos^2\varphi_v \\
    && + \left( \frac{(B_L-B_R)^2(\Delta_L + \Delta_R)^2 + 4 t_c^2 \sin^2 \varphi_v}{B_L-B_R+\Delta_L+\Delta_R} \right)^2\Bigg]^\frac{1}{2} \Bigg\}, \nonumber\\
    \hat E_5' &\approx &  \frac{1}{4} \Bigg( B_L - \sqrt{(B_L-B_R)^2 + 32 t_c^2 \cos^2\varphi_v} \nonumber\\
    && + 3B_R + \frac{2(\Delta_L - \Delta_R)^2}{\sqrt{(B_L-B_R)^2 + 32 t_c^2 \cos^2 \varphi_v}} \Bigg), \\
    \hat E_6' &\approx & \frac{B_L + B_R}{2} + \frac{(\Delta_L - \Delta_R)^2}{4 \sqrt{(B_L - B_R)^2 + 32 t_c^2 \cos^2 \varphi_v}}. \label{eq_E6p}
\end{eqnarray}
Here, again one finds $\hat E_5' \approx \hat E_6' - \omega_p$.

\begin{figure}
\begin{center}
\includegraphics[width=0.5\textwidth]{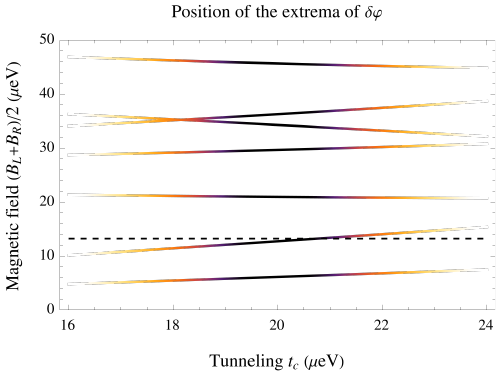}
\caption{Center positions of the dips of $\delta\varphi$ in Fig.~\ref{fig_valleyChar}(a) as a function of $t_c$ and $B_L+B_R$. The color of the lines indicates their normalized visibility, the darker the curve the more prominent the dip. The transitions that are used for the characterization (solid) become off-resonant upon changing $t_c$ and decrease with a Lorentzian lineshape, while the parasitic resonance (dashed), Eqs.~(\ref{eq_firstOtherResonance}-\ref{eq_lastResonance}), remains at maximum visibility. Since the positions of the ALCs depend on the tunneling the dips shift along the $B_L+B_R$-axis as a function of $t_c$. Both effects together can help identifying the desired dips and distinguishing them from the undesired resonances.\label{fig_resonancePositions}}
\end{center}
\end{figure}

The positions of the level crossings depend on the tunneling $t_c$. This is shown in Fig.~\ref{fig_resonancePositions} exemplary for the seven ALCs that appear in the cavity response shown in Fig.~\ref{fig_valleyChar}(a) and are discussed in the main text in Sec.~\ref{sec_valleyChar}. The resonance condition chosen for the characterization of the DQD, $\omega_p \approx\sqrt{2}|t_c \cos\varphi_v|$, is only satisfied at one specific value of $t_c$. However, the effect of the ALCs on the phase shift $\delta\varphi$ vanishes if $t_c$ is detuned from that value.

\bibliography{blockade+resonator_lit.bib}

\end{document}